\documentclass[aps, preprint, groupedaddress, floatfix]{revtex4}

\usepackage{epsfig}
\usepackage{graphicx}
\usepackage{amsmath}
\usepackage{tabularx}
\usepackage{multirow}
\usepackage{color}
\usepackage{tabularx, ragged2e}
\usepackage{booktabs}

\begin{document}

\title{Design of new Mott multiferroics via complete charge transfer: promising candidates for bulk
  photovoltaics}

\author{Hanghui~Chen$^{1,2}\footnote{Correspondence to hanghui.chen@nyu.edu}$ and Andrew Millis$^{3}$}

\affiliation{
 $^1$NYU-ECNU Institute of Physics, NYU Shanghai, Shanghai 200062, China\\
 $^2$Department of Physics, New York University, New York, NY  10002, USA\\ 
 $^3$Department of Physics, Columbia University, New York, NY, 10027, USA\\
}
\date{\today}

\begin{abstract} 
Optimal materials to induce bulk photovoltaic effects should lack inversion symmetry and have an optical gap matching the energies of visible radiation. Ferroelectric perovskite oxides such as BaTiO$_3$ and PbTiO$_3$ exhibit substantial polarization and stability, but have the disadvantage of excessively large band gaps. We use both density functional theory and dynamical mean field theory calculations to design a new class of Mott multiferroics--double perovskite oxides $A_2$VFeO$_6$ ($A$=Ba, Pb, etc). While neither perovskite $A$VO$_3$ nor $A$FeO$_3$ is ferroelectric, in the double perovskite $A_2$VFeO$_6$ a `complete' charge transfer from V to Fe leads to a non-bulk-like charge configuration--an empty V-$d$ shell and a half-filled Fe-$d$ shell, giving rise to a polarization comparable to that of ferroelectric $A$TiO$_3$. Different from nonmagnetic $A$TiO$_3$, the new double perovskite oxides have an antiferromagnetic ground state and around room temperatures, are paramagnetic Mott insulators. Most importantly, the V $d^0$ state significantly reduces the band gap of $A_2$VFeO$_6$, making it smaller than that of $A$TiO$_3$ and BiFeO$_3$ and rendering the new multiferroics a promising candidate to induce bulk photovoltaic effects.  \end{abstract}

\maketitle

\section{Introduction}

The lack of inversion symmetry caused by ferroelectric ordering in certain transition metal oxides can separate the electrons and holes generated by photo-excitation, making these materials promising candidates for photovoltaic devices~\cite{Yang2010,Cao2012,Alexe2011,Grinberg2013}. However, many known ferroelectric perovskite oxides including BaTiO$_3$ and PbTiO$_3$ have very large band gaps ($\sim$ 3-5 eV)~\cite{Bilc2008}, significantly limiting their absorption efficiency in the visible frequency range. The large band gap is intrinsic: it is set by the energy difference between the Ti-$d$ and O-$p$ levels, which is large because Ti and O have substantially different electronegativity. Intensive research in perovskite oxides has focused on reducing band gaps while maintaining ferroelectric polarization. One approach is to replace a fraction of transition metal ions with a different cation; with one transition metal species driving ferroelectricity and the other providing lower energy states that reduce the band gap~\cite{Bennett2008, Gou2011, Qi2011, Choi2009, Choi2012, Nechache2014}. Using this approach, band gap reductions by $\sim$ 1 eV have been attained~\cite{Choi2012} and a high power conversion efficiency has been experimentally achieved in Bi$_2$FeCrO$_6$~\cite{Nechache2014}. In another method, a class of layered double perovskite oxides $AA'BB'$O$_6$ has been theoretically proposed, in which a large in-plane polarization is found via nominal $d^0$ filling on the $B$-site, $A$-site cations bearing lone-pair electrons, and $A'\neq A$ size mismatch; the band gap is controlled by $B/B'$ electronegativity difference~\cite{Kim2015}.

In this work, we propose a simple design scheme. We introduce a new class of double perovskite oxides $A_2$VFeO$_6$ where $A$ is a divalent cation ($A$=Ba, Pb, etc) and demonstrate that a `complete' charge transfer (nominally one electron transfer) between the two transition metal ions~\cite{Chen2013, Chen2013a, Kleibeuker2014, Disa2015, Cao2016, Grisolia2016} can induce desirable properties for bulk photovoltaics. First-principles calculations show that while neither bulk perovskite $A$VO$_3$ nor $A$FeO$_3$ is ferroelectric, a `complete' charge transfer occurs from V to Fe, rendering the new double perovskite oxides a Mott multiferroic: at zero temperature a ferroelectric antiferromagnet and around room temperatures a ferroelectric Mott insulator. The ferroelectric polarization is substantial, comparable to $A$TiO$_3$, but the band gap is significantly lower, smaller than that of $A$TiO$_3$ and BiFeO$_3$.

\begin{figure}[h!]
\includegraphics[angle=0,width=\textwidth]{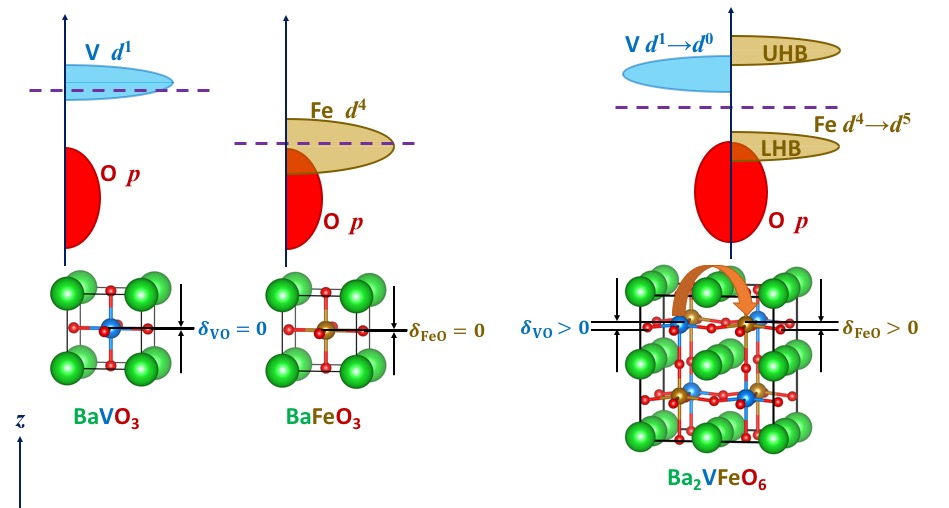}
\caption{\label{fig:schematics} \textbf{Design principles for
    charge-transfer-driven Mott multiferroics.} \textbf{a}) Energy diagram and atomic
  structure of cubic BaVO$_3$. The dashed line is the Fermi
  level. $\delta_{\rm VO}$ is the V-O displacement along the [001]
  direction. \textbf{b}) Energy diagram and atomic structure of cubic
  BaFeO$_3$. The dashed line is the Fermi
  level. $\delta_{\rm FeO}$ is the Fe-O displacement along the [001]
  direction. \textbf{c}) Energy diagram and atomic structure of double perovskite
  Ba$_2$VFeO$_6$. The dashed line is the Fermi level, which lies in
  the gap between V $d$ and Fe $d$ states. `LHB' (`UHB') means lower
  Hubbard bands (upper Hubbard bands). The red arrow indicates the
  charge transfer from V atoms to Fe atoms due to electronegativity
  difference. In the double perovskite
  Ba$_2$VFeO$_6$, a polar distortion is developed
  ($\delta_{\rm VO} > 0$ and $\delta_{\rm FeO} > 0$) because of the new
  charge configuration V $d^0$ and Fe $d^5$.}
\end{figure}

We first focus on Ba$_2$VFeO$_6$ (similar results are obtained for Pb$_2$VFeO$_6$ and Sr$_2$VFeO$_6$, see section~\ref{Discussion}). Fig.~\ref{fig:schematics}\textbf{a} and \textbf{b} show the atomic and electronic structures for perovskite BaVO$_3$ and BaFeO$_3$, respectively. Bulk perovskite BaVO$_3$ has been recently synthesized at high pressure and has been found to remain cubic and metallic to  the lowest temperature~\cite{Nishimura2014}. Bulk BaFeO$_3$ normally crystallizes in a hexagonal structure  but cubic  perovskite BaFeO$_3$ can be stabilized in powders~\cite{Hayashi2011} and in epitaxial thin films~\cite{Matsui2002, Matsui2003, Callender2008, Chakraverty2013} and exhibits a robust ferromagnetism~\cite{Hayashi2011, Callender2008, Matsui2002, Matsui2003, Chakraverty2013}. Both metallic~\cite{Callender2008, Hayashi2011} and insulating~\cite{Matsui2002, Matsui2003, Chakraverty2013} behaviors have been reported.

Formal valence considerations imply that in BaVO$_3$ the V adopts a $d^1$ configuration while in BaFeO$_3$ the Fe is $d^4$. In the double perovskite Ba$_2$VFeO$_6$, however, we expect that the large electronegativity  difference between V and Fe leads to complete charge transfer from V to Fe, resulting in  V-$d^0$ and Fe-$d^5$ configurations as illustrated in Fig.~\ref{fig:schematics}\textbf{c}.  Similar phenomena have been predicted and observed in many different transition metal oxide heterostructures~\cite{Kleibeuker2014, Disa2015, Cao2016, Grisolia2016,Chen2016}. The particular relevance here is that the empty V-$d$ shell and half-filled Fe-$d$ shell are both susceptible to noncentrosymmetric distortions (for the empty $d$ shell case, see~\cite{Eng2003, Berger2012} and for the half-filled $d$ shell cases see~\cite{Neaton2005, Rondinelli2009, Chen2016a}) while Ba$^{2+}$-O$^{2-}$ coupling stabilizes the ferroelectric phase over anti-ferroelectric phases, as in
BaTiO$_3$~\cite{Benedek2016}. The half filled Fe-$d$ shell leads to magnetic ordering and Mott insulating behavior, while the position of the V-$d$ level leads to a reduced band gap (a similar strategy to reduce band gap has been discussed in Refs.~\cite{Eng2003, Berger2012, Kim2015}. Therefore as Fig.~\ref{fig:schematics}\textbf{c} shows, double perovskite Ba$_2$VFeO$_6$ is predicted to be Mott multiferroic (paramagnetic ferroelectric at high temperatures and long-range magnetically ordered at sufficiently low temperatures). Furthermore, as illustrated in
Fig.~\ref{fig:schematics}\textbf{c}, the band gap of double perovskite Ba$_2$VFeO$_6$ is set by the filled lower Hubbard band of Fe-$d$ states (strongly hybridized with O-$p$ states) and empty V-$d$ states (conduction band edge). 

We note that the double perovskite structure is much more stable than the layered configuration as proposed in Ref.~\cite{Kim2015}, because charge transfer generically results in substantial metal-oxygen bond disproportionation~\cite{Chen2016}. Due to geometry consideration, the bond disproportionation inevitably induces internal strain in the layered structure but is naturally accommodated 
by the double perovskite structure, which explains the phase stability~\cite{Chen2016}. Also 
different from previous speculation that rock-salt ordering of $B$-site atoms suppress polarization 
in $A_2BB'$O$_6$~\cite{Knapp2006, Kim2015}, our work shows that it is possible to induce robust ferroelectricity in double perovskite oxides Ba$_2$VFeO$_6$. 

In the rest of this paper we present calculations substantiating this picture. In Section ~\ref{methods} we outline the computational details. In Section~\ref{Results} we present results for double perovskite Ba$_2$VFeO$_6$. Section~\ref{Discussion} extends the calculations to the double perovskite Pb$_2$VFeO$_6$ and Sr$_2$VFeO$_6$, in which we discuss the similarities and differences. Section~\ref{conclusion} is a summary and conclusion. 

\section{Computational Details \label{methods}}

Our first-principles calculations are performed using density functional theory (DFT)~\cite{Payne1992} and dynamical mean field theory (DMFT)~\cite{Kotliar2006}.  Structural relaxation is performed within DFT. Gaps are calculated using both DFT and DFT+DMFT. It is known in literature that structural and magnetic properties of multiferroic oxides strongly depend on the choice of exchange correlation functionals~\cite{Bilc2008, Chen2016a, Chen2016b}. We use three exchange correlation functionals to test the robustness of our predictions: i) charge-density-only generalized gradient approximation with Perdew-Burke-Ernzerhof parametrization~\cite{Perdew1996} plus Hubbard $U$ and Hund's $J$ corrections (PBE+$U$+$J$)~\cite{Liechtenstein1995}, ii) charge-only local density approximation with Hubbard $U$ and Hund's $J$ corrections (LDA+$U$+$J$)~\cite{Kohn1965, Liechtenstein1995}; iii) spin-polarized generalized gradient approximation with Perdew-Burke-Ernzerhof parametrization revised for solids (sPBEsol)~\cite{Perdew2008}. In order to investigate Mottness and effects of long-range magnetic ordering, we use DMFT to study both paramagnetic and long-range magnetic ordered states. 

The DFT calculations are performed using a plane-wave basis~\cite{Payne1992}, as implemented in the Vienna Ab-initio Simulation Package (VASP)~\cite{Kresse1996, Kresse1996a}. The Projector Augmented Wave (PAW) approach is used~\cite{Blochl1994, Kresse1999}.  We use an energy cutoff of 600 eV. All the supercells of double perovskite oxides $A_2$VFeO$_6$ consist of 40 atoms to accommodate different magnetic orderings. We consider ferromagnetic ordering, [001] antiferromagnetic ordering, [010] antiferromagnetic ordering and [100] antiferromagnetic ordering (see the Supplementary Materials for their definitions).  A $6\times 6\times 6$ Monkhorst-Pack grid is used to sample the Brillouin zone. Both cell and internal coordinates are fully relaxed until each force component is smaller than 10 meV/\AA~and the stress tensor is smaller than 0.1 kbar. 

In the PBE+$U$+$J$/LDA+$U$+$J$ as well as DMFT calculations, we use $U_{\textrm{Fe}}$ = 5 eV, $J_{\textrm{V}} = J_{\textrm{Fe}} = 0.7$ eV, following previous studies~\cite{Pavarini2004a, Mosey2008}. The choice of $U_{\textrm{V}}$ needs caution. While $U_{\textrm{V}}$ of about 5 eV has been accepted in literature~\cite{Pavarini2004a}, we find that $U_{\textrm{V}}$ = 5 eV induces off-center displacement $\delta_{\textrm{VO}}$ in perovskite BaVO$_3$, while in experiment the perovskite BaVO$_3$ is stabilized in a cubic structure under 15 GPa~\cite{Nishimura2014}. The off-center displacement of V is closely related to orbital ordering ($d^1_{xy}d^0_{xz}d^0_{yz}$) stabilized by a large $U_{\textrm{V}}$ in the DFT+$U$ method. Therefore we use a smaller $U_{\textrm{V}}$ = 3 eV which stabilizes a cubic structure in perovskite BaVO$_3$ to calculate double perovskite oxides Ba$_2$VFeO$_6$. This ensures that a non-zero $\delta_{\textrm{VO}}$ in Ba$_2$VFeO$_6$ is not a consequence of a large $U_{\textrm{V}}$, but rather is induced by charge transfer. We repeat all the DFT calculations on Ba$_2$VFeO$_6$ using $U_{\textrm{V}}$ = 5 eV and find qualitatively similar results in structural properties. On the other hand, $U_{\textrm{V}}$ controls the energy level of V-$d$ states, which may affect the band gap of Ba$_2$VFeO$_6$. Therefore, in our DMFT calculations, we also study a range of $U_{\textrm{V}}$ (from 3 to 6 eV) to estimate the variation of energy gap in the spectral function.
    
We perform single-site DMFT calculations with Ising-like Slater-Kanamori interactions. The impurity problem is solved using the continuous-time quantum Monte Carlo algorithm with a hybridization expansion~\cite{Werner2006, Gull2011}. The correlated subspace and the orbitals with which it mixes are constructed using maximally localized Wannier functions~\cite{Marzari2012} defined over the full 10 eV range spanned by the $p$-$d$ band complex, resulting in a well-localized set of $d$-like orbitals. All the DMFT calculations are performed at the temperature of 290 K. For each DMFT iteration, a total of 3.8 billion Monte Carlo steps is taken to converge the impurity Green function and self energy. In double perovskite oxides, since V-$d$ states are empty, we treat V-$t_{2g}$ orbitals with the DMFT method and V-$e_g$ orbitals with a static Hartree-Fock approximation. Because Fe-$d$ states are half-filled, we treat all the five Fe-$d$ orbitals with the DMFT method. The two self energies (one for V sites and the other for Fe sites) are solved independently and then coupled at the level of self-consistent conditions.

To obtain the spectral functions, the imaginary axis self energy is continued to the real axis using the maximum entropy method~\cite{Silver1990}.  Then the real axis local Green function is calculated using the Dyson equation and the spectral function is obtained following:

\begin{equation}
\label{eqn:spectral} A_i(\omega) = -\frac{1}{\pi}\textrm{Im} G^{\textrm{loc}}_i(\omega) = -\frac{1}{\pi}\textrm{Im}\left(\sum_{\textbf{k}}\frac{1}{(\omega+\mu)\mathbf{1}-H_0(\textbf{k})-\Sigma(\omega)}\right)_{ii}
\end{equation}
where $i$ is the label of a Wannier function. $\mathbf{1}$ is an identity matrix, $H_0(\textbf{k})$ is the DFT-PBE band Hamiltonian in the matrix form using the Wannier basis. $\Sigma(\omega)$ is understood as a diagonal matrix only with nonzero entries on the correlated orbitals. $\mu$ is the chemical potential.
 $V_{dc}$ is the fully localized limit (FLL) double counting potential, which is defined as~\cite{Czyzyk1994}:

\begin{equation}
\label{eqn:dc} V_{dc} = (U - 2J)\left(N_d - \frac{1}{2}\right) - \frac{1}{2}J(N_d - 3)
\end{equation}
where $N_d$ is the $d$ occupancy of a correlated site.

\section{Results of B\lowercase{a}$_2$VF\lowercase{e}O$_6$ \label{Results}}

\subsection{Structural properties}

We first discuss the fully relaxed atomic structure of double perovskite Ba$_2$VFeO$_6$, obtained using DFT calculations with three different exchange correlation functionals (PBE+$U$+$J$, LDA+$U$+$J$ and sPBEsol). For each exchange correlation functional, we test ferromagnetic ($F$), [001] antiferromagnetic, [010] antiferromagnetic and [100] antiferromagnetic orderings (see the Supplementary Materials for precise definitions).  For each case, we start from a crystal structure with rotations and tilts of VO$_6$ and FeO$_6$ (space group $P2_1/n$) and then perturb the V and Fe atoms along [001] or (011) and (111) directions. Next we perform atomic relaxation with all the symmetry turned off. After atomic relaxation, we find that the rotations and tilts of VO$_6$ and FeO$_6$ are strongly suppressed while the polarization along [001] or (011) or (111) direction is stabilized. Comparing the total energy between three polarizations, we find the ground state of Ba$_2$VFeO$_6$ has the polarization along the [001] direction. The ground state structure has tetragonal symmetry (space group $I4/m$).  On the magnetic properties, given the $U$ and $J$ values, we find that the ground state is always of the [001] antiferromagnetic ordering.
Using the same methods and parameters, perovskite BaVO$_3$ and BaFeO$_3$ have cubic symmetry. The resulting lattice constant $a$, tetragonality $c/a$ ratio and cation-displacement $\delta_{B\textrm{O}}$ along the [001] direction (see in Fig.~\ref{fig:schematics}\textbf{c}) are shown in Table~\ref{tab:Ba-ferroelectric} for each exchange correlation functional. The full crystal structure data are provided in the Supplementary 
Materials. We need to point out that the reason that rotations and tilts of VO$_6$/FeO$_6$ octahedra
are strongly suppressed in Ba$_2$VFeO$_6$ is due to the large ionic size of Ba ions, which is known to prohibit rotations and tilts of oxygen octahedra in perovskite Ba-compounds and to induce robust ferroelectricity in BaTiO$_3$ and BaMnO$_3$~\cite{Wu2011, Rondinelli2009}.

For comparison, we also calculate the atomic structure of fully relaxed tetragonal BaTiO$_3$, a known ferroelectric perovskite. Since BaTiO$_3$ is a $d^0$ band insulator with no magnetic properties, we do not add Hubbard $U$ and Hund's $J$ correction to PBE/LDA and we use PBEsol instead of spin-polarized PBEsol (sPBEsol). We find that the calculated $c/a$ ratio and ion-displacement ($\delta_{\textrm{VO}}$ and $\delta_{\textrm{FeO}}$) of Ba$_2$VFeO$_6$ are comparable to those of BaTiO$_3$. The ground state of 
tetragonal double perovskite Ba$_2$VFeO$_6$ is an insulator (we will discuss the gap properties in details in the following subsections). The ground state of high-symmetry cubic double perovskite Ba$_2$VFeO$_6$ 
is also an insulator (see Table~\ref{tab:Ba-ferroelectric}). Therefore a switching path for ferroelectric polarization is well-defined and we can use the Berry phase method~\cite{Marzari2012} to calculate the polarization of the tetragonal structure. We find that for each exchange-correlation function the calculated polarization of Ba$_2$VFeO$_6$ is comparable to that of BaTiO$_3$ (see Table~\ref{tab:Ba-ferroelectric}).

We comment here that our recent study~\cite{Chen2016a, Chen2016b} of perovskite manganites show that PBE+$U$+$J$ and sPBEsol yield the most accurate predictions on structural and magnetic properties of magnetic ferroelectrics, while LDA+$U$+$J$ sets an conservative estimation for the lower bound of polarization. Therefore we believe that the polarization of Ba$_2$VFeO$_6$ is larger than 18 $\mu$C/cm$^2$, which is substantial enough to induce bulk photovoltaic effects~\cite{Grinberg2013}.

\subsection{Electronic properties}

The results of the previous subsection indicate that the double perovskite Ba$_2$VFeO$_6$ has a noncentrosymmetric tetragonal distortion not found in the component materials bulk BaVO$_3$ and BaFeO$_3$. In this section we consider the electronic reconstruction arising in the double perovskite.

\begin{figure}[h!]
\includegraphics[angle=0,width=\textwidth]{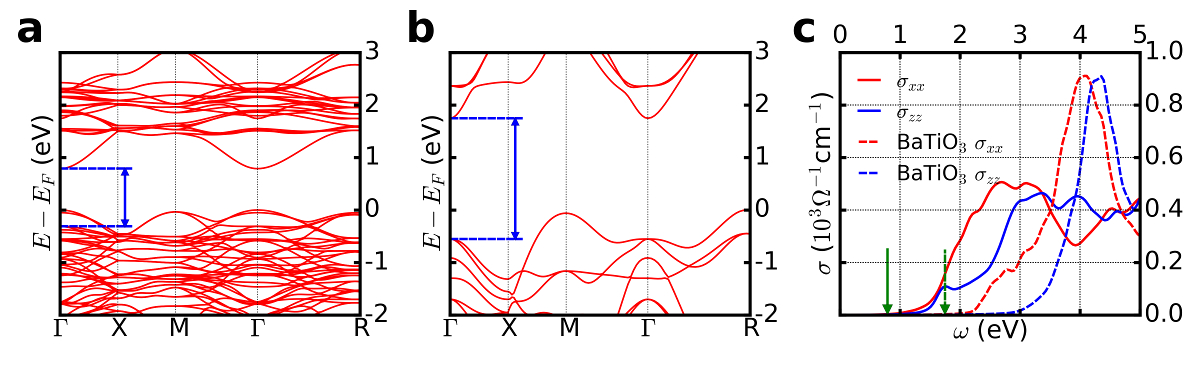}
\caption{\label{fig:optical} \textbf{Comparison of band structure and
    optical conductivity between Ba$_2$VFeO$_6$ and BaTiO$_3$.} 
  The results for Ba$_2$VFeO$_6$ are calculated using DFT-PBE+$U$+$J$
  method. The results for BaTiO$_3$ are calculated using DFT-PBE method.
  \textbf{a}) Band structure of tetragonal
  Ba$_2$VFeO$_6$. The blue arrow indicates the threshold of optical
  transition. \textbf{b}) Band structure of tetragonal
  BaTiO$_3$. The blue arrow indicates the threshold of optical
  transition. \textbf{c}) Optical conductivity $\sigma$ of
  tetragonal Ba$_2$VFeO$_6$ (solid lines) and tetragonal BaTiO$_3$
  (dashed lines). The red lines are for the $xx$-component and the
  blue lines are for the $zz$-component. The green arrows indicate the
  fundamental gap of band structures.}
\end{figure}

Fig.~\ref{fig:optical}\textbf{a} shows the band structure of double perovskite Ba$_2$VFeO$_6$ with the [001] antiferromagnetic ordering (only one spin channel is shown here), calculated using the PBE+$U$+$J$ method. We see that a gap is clearly opened in Ba$_2$VFeO$_6$ while using the same method with the same parameters, perovskite BaVO$_3$ and BaFeO$_3$ are found to be metallic with V-$d$ and Fe-$d$ states at the Fermi surface (see Section II in the Supplementary Materials for details).  The gap opening in Ba$_2$VFeO$_6$ is a strong evidence of a nominally ``complete'' charge transfer from V to Fe. A similar charge-transfer-driven metal-insulator transition is predicted~\cite{Chen2013b} and observed~\cite{Cao2016} in LaTiO$_3$/LaNiO$_3$ superlattices.

For comparison, we also calculate the band structure of tetragonal BaTiO$_3$ using PBE (Fig.~\ref{fig:optical}\textbf{b}). We note that while the polarization of double perovskite Ba$_2$VFeO$_6$ is comparable to that of BaTiO$_3$, the band gap of Ba$_2$VFeO$_6$ (0.78 eV) is significantly smaller than that of BaTiO$_3$ (1.75 eV). Using other exchange correlation functionals, we find similar properties that the band gap of Ba$_2$VFeO$_6$ is smaller than that of BaTiO$_3$ by about 1 eV (see `fundamental gap' $\Delta_0$ in Table~\ref{tab:Ba-ferroelectric}).

For photovoltaic effects the relevant quantity is the optical gap $\Delta_{\textrm{optical}}$. We calculate the optical conductivity of both Ba$_2$VFeO$_6$ and BaTiO$_3$ using standard methods ~\cite{Gajdos2006} and show the results in Fig.~\ref{fig:optical}\textbf{c}.  Due to the tetragonal symmetry, the off-diagonal matrix elements of the optical conductivity vanish and only two diagonal elements are independent ($\sigma_{xx}=\sigma_{yy}$ and $\sigma_{zz}$).  For BaTiO$_3$ the minimum optical gap is in the $xx$ channel and is given by the direct (vertical in momentum space) gap (shown for BaTiO$_3$ as the blue arrow in Fig.~\ref{fig:optical}\textbf{b}). In BaTiO$_3$ the optical gap is larger than the fundamental gap, which is indirect (momentum of lowest conduction band state differs from momentum of highest valence band state; the green arrow in Fig.~\ref{fig:optical}\textbf{c} shows the size of the fundamental gap).
The optical conductivity of Ba$_2$VFeO$_6$ is also larger than its fundamental gap, which can be understood in a similar manner. If we  consider (VFe) as a pseudo-atom $X$, the hypothetical single perovskite Ba$X$O$_3$ would have an indirect gap (between $\Gamma$ and $R$). However, the reduction in translational symmetry due to the V-Fe alternation leads to  band folding which maps the original $R$ point to the $\Gamma$ point, leading to a direct gap of 0.8
eV  at the $\Gamma$ point. However the calculated optical gap is 1.1 eV (blue arrow in Fig.~\ref{fig:optical}\textbf{a}).  The difference between the direct and optical gaps is a matrix element effect: the lowest back-folded conduction band state does not have a dipole allowed transition matrix element with the highest-lying valence band state (see the Supplementary Materials for more details). 

It is well-known that DFT with semi-local exchange correlation functionals substantially underestimate band gaps. Here we argue that since Ba$_2$VFeO$_6$ and BaTiO$_3$ have similar electronic structures (gap separated by metal $d$ and oxygen $p$ states), the DFT band gap underestimation with respect to experimental values is approximately a constant for BaTiO$_3$ and Ba$_2$VFeO$_6$. The experimental optical gap of BaTiO$_3$ is 3.2 eV and the DFT calculated value is 2.3 eV, about 0.9 eV smaller. The DFT calculated optical gap of Ba$_2$VFeO$_6$ is 1.1 eV, hence we estimate the experimental optical gap of Ba$_2$VFeO$_6$ is 2.0 eV, which is smaller than the optical gap of intensively investigated BiFeO$_3$ (2.7 eV)~\cite{Ihlefeld2008}.

\begin{figure}[h!]
\includegraphics[angle=0,width=0.85\textwidth]{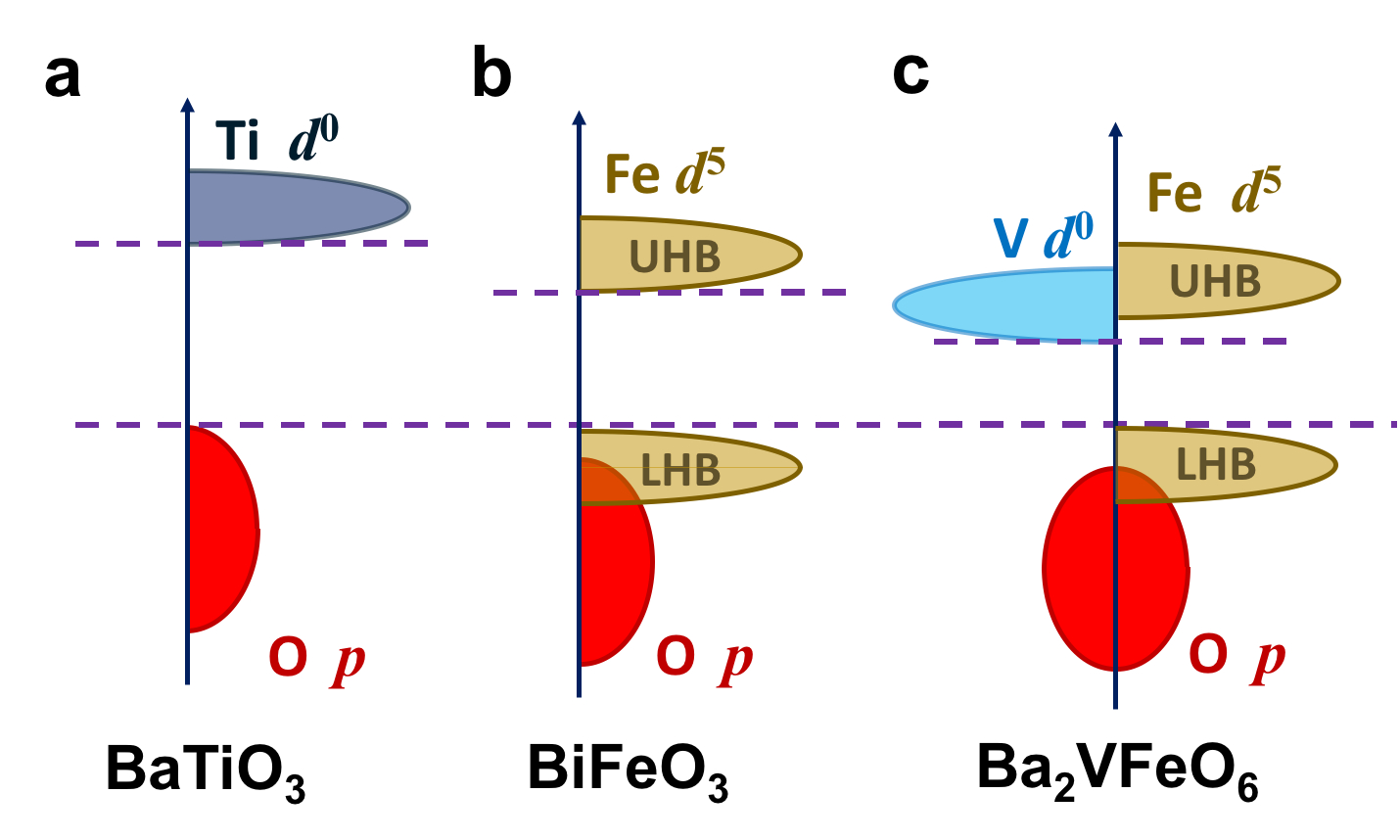}
\caption{\label{fig:estimation} \textbf{Comparison of gaps for different perovskite oxides.} \textbf{a}) BaTiO$_3$; \textbf{b}) BiFeO$_3$; \textbf{c}) Ba$_2$VFeO$_6$. `LHB' (`UHB') means lower
  Hubbard bands (upper Hubbard bands). The valence band edges are aligned for comparison.}
\end{figure}

We comment here that while we use the assumption that our DFT band gap underestimation (0.9 eV) applies to both BaTiO$_3$ and Ba$_2$VFeO$_6$, our results that Ba$_2$VFeO$_6$ should have a smaller gap than that of BaTiO$_3$ and BiFeO$_3$ are supported by physical arguments (see Fig.~\ref{fig:estimation}). The band gap for transition metal oxides is set by the energy difference between transition metal $d$ states and oxygen $p$ states. This $p$-$d$ separation is a measure of the relative electronegativity of transition metal and oxygen ions. Ti and V are both first-row transition metals and in BaTiO$_3$ and Ba$_2$VFeO$_6$, Ti and V both have a $d^0$ configuration. Because V has a larger nuclear charge than Ti, the V-$d$ states have lower energies than the Ti-$d$ states, which leads to a smaller band gap for Ba$_2$VFeO$_6$ than for BaTiO$_3$ (compare panels \textbf{a} and \textbf{c} of Fig.~\ref{fig:estimation}). On the other hand, the Fe $d$ states are half-filled in both Ba$_2$VFeO$_6$ and BiFeO$_3$, while V-$d$ states are empty in Ba$_2$VFeO$_6$. Due to Coulomb repulsion and Hund's coupling effects, adding one more electron in a half-filled $d$ shell generically costs more energy than adding an electron in an empty $d$ shell. Therefore the upper Hubbard band of Fe $d$ states have higher energy than V $d$ states, which results in a larger band gap for BiFeO$_3$ than for Ba$_2$VFeO$_6$ (compare panels \textbf{b} and \textbf{c} of Fig.~\ref{fig:estimation}).   

\subsection{Estimation of critical temperatures}

Double perovskite Ba$_2$VFeO$_6$ is a  type-I multiferroic~\cite{Khomskii2009}, in which ferroelectric polarization and magnetism arise from different origins and they are largely independent of one another. This means that ferroelectric polarization and magnetism have their own critical temperatures and usually the critical temperature of polarization ($T_C$) is higher than the critical temperature of magnetism ($T_N$)~\cite{Puggioni2015}. In this subsection, we estimate $T_C$ and $T_N$ for Ba$_2$VFeO$_6$. 

\textit{Estimation of $T_C$}: in order to estimate the ferroelectric Curie temperature $T_C$, we use the predictor $T_C \propto P^2_0$ where $P_0$ is the zero-temperature polarization~\cite{Abrahams1968}. This predictor has been successfully applied to a wide range of Pb-based perovskite ferroelectric oxides and it yields an accurate and quantitative estimation for ferroelectric $T_C$~\cite{Grinberg2004}. We apply this predictor to our Ba-based ferroelectrics, i.e. BaTiO$_3$ and Ba$_2$VFeO$_6$. Here we use tetragonal BaTiO$_3$ as the reference system. The experimental Curie temperature $T_C$ for BaTiO$_3$ is about 400 K~\cite{Megaw1946}. Using the DFT+Berry phase method~\cite{Marzari2012}, we can obtain the values of the zero-temperature polarization for both BaTiO$_3$ and Ba$_2$VFeO$_6$ shown in Table~\ref{tab:Ba-ferroelectric}. Therefore we estimate that $T_C$ for Ba$_2$VFeO$_6$ is 473 K (PBE+$U$+$J$), 245 K (LDA+$U$+$J$) and 425 K (sPBEsol). While different exchange correlation functionals predict a range for $T_C$, we find that $T_C$ is near or above room temperature. 

%The other predictor is
%to use a Heisenburg-like Hamiltonian $E=\frac{1}{2}J\sum_{\langle ij
%\rangle}\mathbf{P}_i\cdot\mathbf{P}_j$ where $\textbf{P}_i$ is the
%polarization at site $i$ (the magnitude is set to unit). We stabilize
%an antiferroelectric state (see Supplementary Materials) and extract
%that $J$ = 8.6 meV (PBE+$U$+$J$), xxx and xxx. Therefore $T_C=kJ=6J$=
%599 K (PBE+$U$+$J$), xx and xxx. where $k$ is the
%number of nearest numbers.  

\textit{Estimation of $T_N$}: we use a classical Heisenberg model $E=\frac{1}{2}\sum_{\langle kl \rangle}J_{kl}\mathbf{S}_k\cdot\mathbf{S}_l$ to estimate the magnetic ordering transition temperature $T_N$, where $\textbf{S}_k$ is a unit-length classical spin and $\langle kl \rangle$ denotes summation over nearest Fe neighbors. Here we only consider Fe-Fe exchange couplings.  Because double perovskite Ba$_2$VFeO$_6$ has a tetragonal structure, there are two exchange couplings of $J_{kl}$: $J_{\textrm{in}}$ for the short Fe-Fe bonds and $J_{\textrm{out}}$ for the long Fe-Fe bonds. By calculating the total energy for the ferromagnetic ordering, [001] antiferromagnetic ordering and [100] antiferromagnetic ordering, we obtain that the in-plane exchange coupling $J_{\textrm{in}}$ is 2.5 meV (PBE+$U$+$J$), 3.7 meV (LDA+$U$+$J$) and 3.1 meV (sPBEsol); and the out-of-plane exchange coupling $J_{\textrm{out}}$ is 3.1 meV (PBE+$U$+$J$), 4.0 meV (LDA+$U$+$J$) and 3.7 meV (sPBEsol). The positive sign means that exchange couplings are all 
antiferromagnetic. Based on a mean-field theory, the estimated N\'{e}el temperature is $T_N=|4J_{\textrm{in}} - 8J_{\textrm{out}}|$. The minus sign is because on a quasi face-centered-cubic lattice, every Fe atom has 
8 nearest neighbors that are antiferromagnetically coupled and 4 nearest neighbors that are ferromagnetically coupled. Therefore $T_N$ is estimated to be 172 K (PBE+$U$+$J$), 200 K (LDA+$U$+$J$) and 200 K (sPBEsol). Since mean-field theories usually overestimate magnetic 
transition temperatures, the actual $T_N$ could be lower.
An experimental determination of the magnetic ordering temperature would be of great interest.

%While all the exchange correlation functionals predict that $T_N$ exceeds the room temperature, these are the mean-field predictions, which usually overestimate magnetic critical temperatures. We used the same method to calculate the N\'{e}el temperature of cubic SrMnO$_3$ finding that while the experimental $T_N$ is 260K, the estimated mean-field value is about 1400 K, larger by a factor of 5. While we do not claim that this reduction factor is universal for all materials, the reduction effect implies that experimental $T_N$ for Ba$_2$VFeO$_6$ should be below the room temperature. We note that the Curie temperature of bulk perovskite BaFeO$_3$ is 235K~\cite{Callender2008}, which is already below the room temperature. In the double perovskite Ba$_2$VFeO$_6$, the nearest Fe sites are further separated by V sites, which suggests that the magnetic transition temperature may be lower than that of BaFeO$_3$. An experimental determination of the magnetic ordering temperature would be of great interest.

\subsection{Effects of long-range orders}

The estimates for the ferroelectric and magnetic transition temperatures of Ba$_2$VFeO$_6$ suggest that its actual ferroelectric Curie temperature $T_C$ is probably higher than its actual N\'{e}el temperature $T_N$, as is the case for most type-I multiferroics~\cite{Khomskii2009}. It is therefore important to ask if the magnetically disordered state remains insulating, so that the ferroelectric properties are preserved.

\begin{figure}[h!]
\includegraphics[angle=0,width=\textwidth]{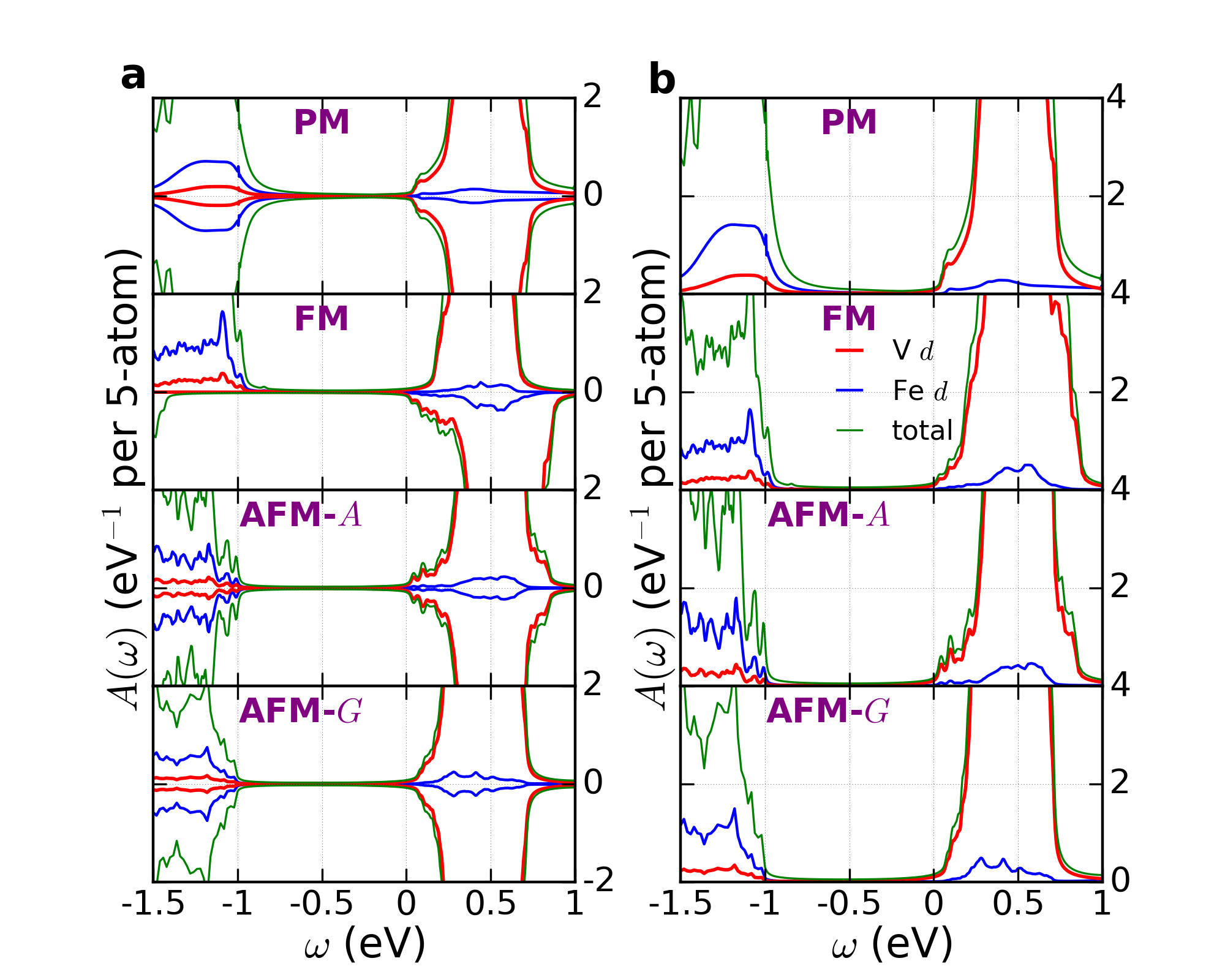}
\caption{\label{fig:mag} \textbf{Spectral functions $A(\omega)$ of tetragonal double
  perovskite Ba$_2$VFeO$_6$ for different magnetic states.} The unit
  of $A(\omega)$ is eV$^{-1}$ per 5-atom.
  `PM' stands for paramagnetic state, `FM' for ferromagnetic state,
  `[001]-AFM' for [001] antiferromagnetic state and `[100]-AFM' for 
  [100] antiferromagnetic state. Panels \textbf{a})
  spin-resolved spectral function. The positive (negative) $y$-axis corresponds
  to spin-up (spin-down). Panels \textbf{b}) total spectral functions (summing
  over spin-up and spin-down). The red, blue and green curves are for
  Fe $d$, V $d$ and O $p$, respectively. The Fermi level is set
  at $\omega=0$ eV.}
\end{figure}

Here we use DFT+DMFT to study both the paramagnetic and magnetically ordered states. The spectral functions for the three magnetic states that we have considered are shown in Fig.~\ref{fig:mag} along with the spectral function for the paramagnetic state. We find that the paramagnetic state is insulating, with a gap only slightly smaller than that of the ground state with [001] antiferromagnetic ordering, indicating that double perovskite Ba$_2$VFeO$_6$ is a promising candidate for Mott multiferroics~\cite{Puggioni2015}. The calculated spectral functions are consistent with our schematics of Fig.~\ref{fig:estimation}.

\begin{figure}[h!]
\includegraphics[angle=0,width=8cm]{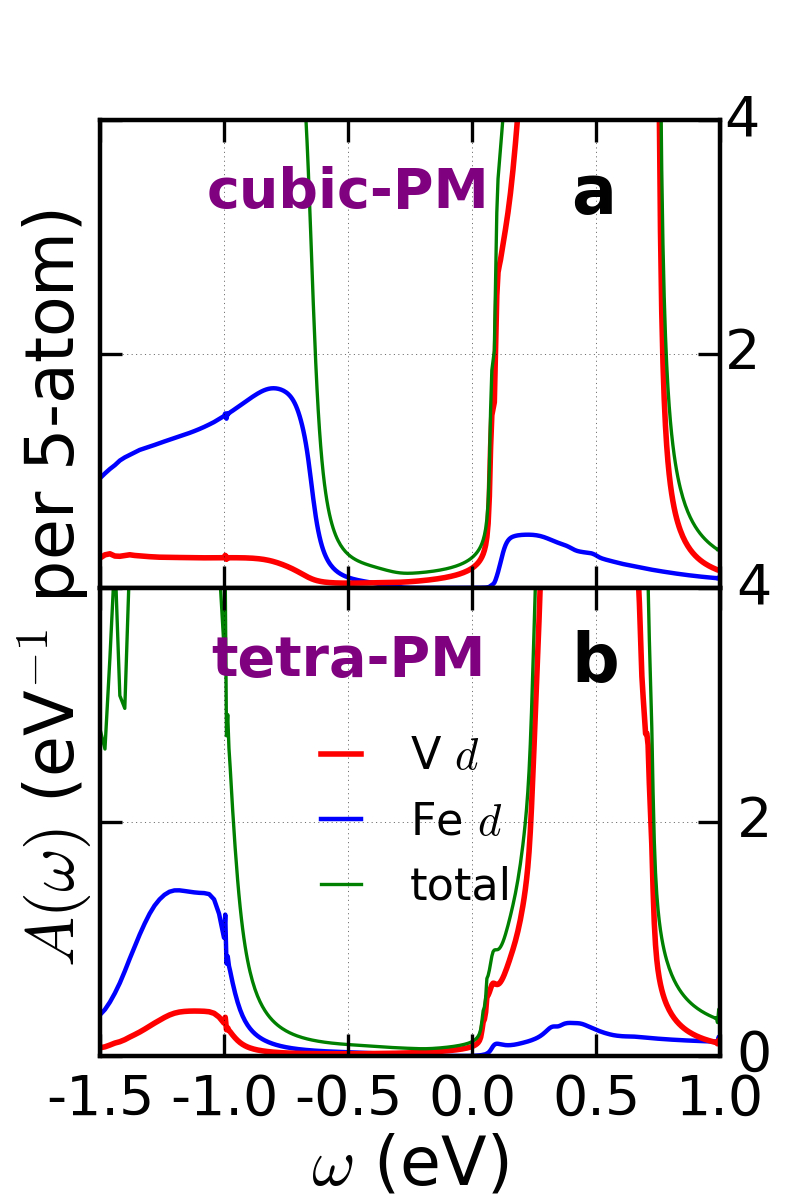}
\caption{\label{fig:cubic} \textbf{Spectral functions $A(\omega)$ of
    cubic and tetragonal Ba$_2$VFeO$_6$.} The unit of $A(\omega)$ is
  eV$^{-1}$ per 5-atom. Panel \textbf{a} is for cubic Ba$_2$VFeO$_6$ and panel \textbf{b} is for
  tetragonal Ba$_2$VFeO$_6$. In both structures, we calculate the
  paramagnetic state. The red, blue and green curves are for
  Fe $d$, V $d$ and O $p$, respectively. The Fermi level is set
  at $\omega=0$ eV.}
\end{figure}

We also use our DFT+DMFT methodology to  investigate how the electronic structure of Ba$_2$VFeO$_6$ evolves as the ferroelectric polarization is suppressed within the paramagnetic state.  Fig.~\ref{fig:cubic} compares the spectral function of Ba$_2$VFeO$_6$ in the cubic structure (i.e no polarization) versus in the tetragonal structure (i.e. with polarization). We see that the suppression of polarization reduces the gap by about 0.2 eV. This behavior is very consistent with similar calculations on nonmagnetic perovskite oxide SrTiO$_3$ in which the presence of ferroelectric polarization can increase the band gap by up to 0.2 eV~\cite{Berger2011}.

\subsection{Hubbard $U$ dependence}

\begin{figure}[h!]
\includegraphics[angle=0,width=\textwidth]{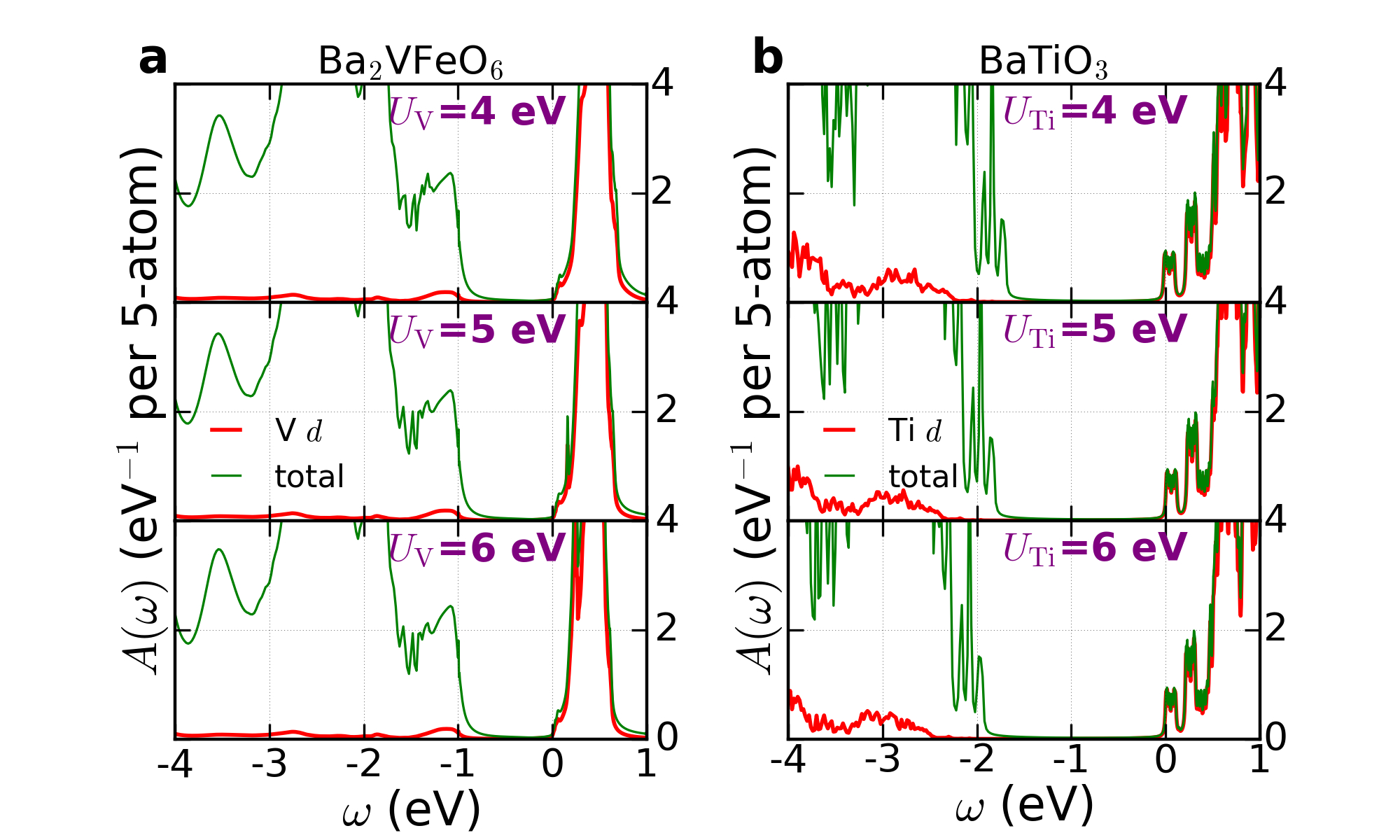}
\caption{\label{fig:gap} \textbf{Spectral functions $A(\omega)$ of tetragonal
  Ba$_2$VFeO$_6$ and BaTiO$_3$
  as a function of Hubbard $U$ on V and Ti.} The unit of $A(\omega)$
  is eV$^{-1}$ per 5-atom. Panels \textbf{a} are the results for tetragonal
  Ba$_2$VFeO$_6$. Panels \textbf{b} are the results for tetragonal BaTiO$_3$.
  For Ba$_2$VFeO$_6$, the calculations are performed in a paramagnetic state. For BaTiO$_3$,
  the calculations are performed in a non-magnetic state. In panels \textbf{a}, the
  green lines are the total spectral functions and the red lines are
  the spectral functions projected onto V $d$ states. In panels \textbf{b},  the
  green lines are the total spectral functions and the red lines are
  the spectral functions projected onto Ti $d$ states. The Fermi
  level is set at $\omega$ = 0 eV.}
\end{figure}

Finally we discuss the Hubbard $U$ dependence. As Fig.~\ref{fig:mag} shows, the conduction band edge is set by V-$d$ states, which is consistent with Fig.~\ref{fig:schematics}\textbf{c} and our previous discussion of band gaps. If we change the Hubbard $U_{\textrm{V}}$, it may affect the energy position of V $d$ states 
and energy gap. To address this issue, we repeat the DMFT calculations on tetragonal Ba$_2$VFeO$_6$ 
using several values of $U_{\textrm{V}}$. The panels \textbf{a} of Fig.~\ref{fig:gap} show the spectral function of the double perovskite as a function of $U_{\textrm{V}}$. All the calculations are performed in a paramagnetic state. We note that as $U_{\textrm{V}}$ increases from 4 eV to 6 eV, the band gap is almost unchanged. This 
is due to the fully localized limit double counting correction which nearly cancels the Hartree shift. Hence, 
the V-$d$ and O-$p$ energy separation is practically unaffected, which is very consistent with the
previous DMFT study on SrVO$_3$~\cite{Dang2014}. If we apply the same method and same Hubbard $U$ parameters to tetragonal BaTiO$_3$, the spectral functions of BaTiO$_3$ (panels \textbf{b} of Fig.~\ref{fig:gap}) show that the energy gap of BaTiO$_3$ is slightly increased.
Thus while we have some uncertainty relating to the appropriate values for the Hubbard $U$, 
our estimates for energy gap are robust: double perovskite Ba$_2$VFeO$_6$ has an energy gap $\sim$ 1 eV smaller than that of BaTiO$_3$. The underlying reason is the differing electronegativities of Ti$^{4+}$ and V$^{5+}$.

\section{Related materials P\lowercase{b}$_2$VF\lowercase{e}O$_6$ and S\lowercase{r}$_2$VF\lowercase{e}O$_6$ \label{Discussion}}

In this section we employ the same parameters and methods used for Ba$_2$VFeO$_6$ to discuss double perovskite Pb$_2$VFeO$_6$ and Sr$_2$VFeO$_6$. 

We first discuss Pb$_2$VFeO$_6$. Pb has a lone pair of $6s$ electrons, which favors off-center displacements as was already shown for tetragonal PbTiO$_3$~\cite{Waghmare1997}. Due to the same mechanism, double perovskite Pb$_2$VFeO$_6$ has substantial cation-displacements, tetragonality and ferroelectric polarization (see Table~\ref{tab:Pb-ferroelectric}). All these values are comparable to, or even larger than those of tetragonal PbTiO$_3$. We note however that within sPBEsol the polarization of this tetragonal structure is not-well defined because the corresponding high-symmetry cubic structure is metallic and thus the obvious switching path is not available. 

While tetragonal double perovskite Pb$_2$VFeO$_6$ have similar structural properties to tetragonal PbTiO$_3$, the fundamental gap $\Delta_0$ and optical gap $\Delta_{\textrm{optical}}$ are both smaller than those of PbTiO$_3$ by about 1 eV (all three exchange correlation functionals make qualitatively consistent predictions).

We note here that the polarization in Pb$_2$VFeO$_6$ has different origin from the polarization in tetragonal PbVO$_3$~\cite{Shpanchenko2004}. In tetragonal PbVO$_3$, V atoms have a $d^1$ charge configuration and its off-center displacement $\delta_{\textrm{VO}}$ and insulating properties are associated with orbital ordering ($d^{1}_{xy}d^0_{xz}d^0_{yz}$)~\cite{Oka2008}. In double perovskite oxide Pb$_2$VFeO$_6$, charge transfer leads to a $d^0$ configuration on V sites and therefore the off-center displacement $\delta_{\textrm{VO}}$ is due to hybridization between V-$d$ and O-$p$ states~\cite{Benedek2016}. More importantly, perovskite
PbVO$_3$ is not ferroelectric because along the switching path (from the tetragonal-to-cubic structure) an insulator-to-metal phase transition is observed~\cite{Belik2005}. 

Next we discuss Sr$_2$VFeO$_6$. Sr$_2$VFeO$_6$ is more complicated because the ionic size of Sr$^{2+}$ is smaller than Ba$^{2+}$ and therefore rotations of oxygen octahedra (so-called antiferrodistortive mode, or AFD mode) can exist in Sr-compounds, such as in SrTiO$_3$, which competes against ferroelectric polarization~\cite{Fleury1968}. For double perovskite Sr$_2$VFeO$_6$, even if we do not take the AFD mode into account, different exchange correlation functionals predict different structural and electronic
properties. Table~\ref{tab:Sr-ferroelectric} shows that PBE+$U$+$J$ predicts that the ground state is tetragonal and ferroelectric. The polarization is sizable (26 $\mu$C/cm$^2$) and the DFT-calculated optical gap is 1.36 eV. On the other hand, the LDA+$U$+$J$ method can not stabilize the tetragonal structure. This method predicts that
ground state of Sr$_2$VFeO$_6$  has a cubic structure with no off-center displacements of either V or Fe, and is metallic. The sPBEsol method can stabilize a tetragonal structure with non-zero off-center displacements $\delta_{\textrm{VO}}$ and $\delta_{\textrm{FeO}}$, but the ground state is also metallic and therefore the polarization is not well-defined. We may impose epitaxial strain to induce ferroelectricity in Sr$_2$VFeO$_6$, but the critical strain strongly depends on the choice of exchange correlation functional~\cite{Chen2016a}: PBE+$U$+$J$ does not require any strain to stabilize the ferroelectric state, while LDA+$U$+$J$ requires a 3\% compressive
strain to open the gap and stabilize the tetragonal structure with a sizable polarization. A similar situation occurs for SrTiO$_3$. If we use the same methods and do not take into account the AFD mode, PBE predicts a ferroelectric ground state, while LDA and sPBE predict that the ground state is cubic (i.e. no polarization). Experimentally, SrTiO$_3$ is on the verge of a paraelectric-to-ferroelectric transition~\cite{Muller1979}. Thus we conclude that our DFT calculations indicate that double perovskite Sr$_2$VFeO$_6$ is close to the paraelectric-to-ferroelectric phase boundary and probably is on the paraelectric side.  

\section{Conclusions~\label{conclusion}}

In summary, we use first-principles calculations to design a new class of Mott multiferroics among which double perovskite oxide Ba$_2$VFeO$_6$ stands out as a promising candidate to induce bulk photovoltaic effects because of its large polarization (comparable to BaTiO$_3$); its reduced optical gap (smaller than BaTiO$_3$ by about 1 eV); and its environmentally friendly composition (Pb-free). Our work shows that charge transfer is a powerful approach to engineering atomic, electronic and magnetic structures in complex oxides. New charge configurations not found in bulk materials can occur in oxide heterostructures (including complex bulk forms such as double perovskites), and these charge configurations can produce emergent phenomena and properties not exhibited in constituent compounds. In particular, V$^{5+}$ is very rare in single perovskite oxides (probably due to its small ionic size). We hope our theoretical predictions can stimulate further experimental endeavors to synthesize and measure these new multiferroic materials for photovoltaic applications.

\begin{acknowledgments}
H. Chen is supported by National Science Foundation
under grant No. DMR-1120296. A. J. Millis is supported by National Science Foundation under grant
No. DMR-1308236. Computational facilities are
provided via Extreme Science and Engineering Discovery
Environment (XSEDE), through Award No. TGPHY130003
and via the National Energy Research Scientific Computing
Center (NERSC).
\end{acknowledgments}

\hspace{-0.5cm}\textbf{Author Contributions}

H. Chen conceived the project and performed numerical calculations. A.J.M supervised the project. 
H. Chen and A.J.M analyzed the data, discussed the results and wrote the manuscript.

\vspace{1cm}

\hspace{-0.5cm}\textbf{Additional Information}

\textbf{Competing Interests:} The authors declare that they have no competing interests.

\clearpage
\newpage

\newpage
\clearpage

\begin{table}[h!]
  \caption{\label{tab:Ba-ferroelectric} \textbf{Comparison of Ba$_2$VFeO$_6$ and
    BaTiO$_3$.} The results are calculated using the DFT method with 
    different exchange correlation functionals (xc).
    `nm' stands for non-magnetic and `[001]' for [001]
    antiferromagnetic ordering. For the cubic case, $a$ is the
    lattice constant and $\Delta_0$ is the fundamental gap. For the tetragonal case, $a$ is the in-plane
    lattice constant, $c/a$ is the ratio of out-of-plane lattice
    constant over in-plane lattice constant,
    $\delta_{\textrm{\textit{B}O}}$ is the $B$-site metal and oxygen displacement along the
    [001] direction. $\Delta_0$ is the fundamental gap and
    $\Delta_{\textrm{optical}}$ is the optical gap. $\Delta E$
    is the energy difference between the tetragonal structure and the
    cubic structure in the unit of meV per 5-atom formula. $P$ is the
    polarization along the [001] direction. $m$ is the local magnetic moment
    on V and Fe sites.}
\begin{center}
\begin{tabularx}{\textwidth}{c| *3{>{\Centering}X}| *3{>{\Centering}X}}
\hline \hline
  & \multicolumn{3}{c|}{Ba$_2$VFeO$_6$}  &  \multicolumn{3}{c}{BaTiO$_3$}  \\
\hline xc &  PBE+$U$+$J$ & LDA+$U$+$J$ & sPBEsol & PBE & LDA & sPBEsol \\
\hline magnetic & [001] & [001] & [001] & nm & nm & nm \\
\hline & \multicolumn{6}{c}{cubic structure} \\
\hline $a$ (\AA) &   4.016  &  3.922 & 3.965 &  4.036  & 3.952  & 3.991 \\
\hline $\Delta_0$ (eV)  & 0.55 & 0.35 & 0.45 & 1.70 & 1.70 & 1.80\\
\hline & \multicolumn{6}{c}{tetragonal structure} \\
\hline $a$ (\AA) & 3.958 & 3.916 & 3.946 & 4.001 & 3.944 & 3.978 \\
\hline $c/a$  & 1.078 & 1.007 & 1.024 & 1.053 & 1.011 & 1.021 \\
\hline $\delta_{B\textrm{O}}$ (\AA) & 0.195 (V)
                                                           0.265 (Fe)
                      & 0.067 (V) 0.086 (Fe) & 0.116 (V) 0.152 (Fe) &
                                                                      0.197 & 0.099 & 0.133 \\
\hline $P$ ($\mu$C/cm$^2$) & 50 & 18 & 34 & 46 & 23  & 33 \\
\hline $\Delta_0$ (eV) & 0.78 & 0.38 & 0.59 & 1.75 & 1.75 & 1.75 \\
\hline $\Delta_{\textrm{optical}}$ (eV) & 1.10 & 1.04 & 1.17 & 2.30 & 2.02
                                                 &  2.14 \\
\hline $\Delta E$ (meV) & -43 & -1 & -7 & -56 & -6 & -17 \\
\hline $m$ ($\mu_B$) & 0.129 (V) 4.023 (Fe) & 0.071 (V)
                                4.075 (Fe)&
                                                                  0.091
                                                                  (V)
 4.063 (Fe) & -- & -- & --\\
\hline\hline
\end{tabularx}
\end{center}
\end{table}

\newpage
\clearpage

\newpage
\clearpage

\begin{table}[h!]
  \caption{\label{tab:Pb-ferroelectric} 
\textbf{Comparison of Pb$_2$VFeO$_6$ and
    PbTiO$_3$.} The results are calculated using the DFT method with 
    different exchange correlation functionals (xc).
    `nm' stands for non-magnetic and `[001]' for the [001]
    antiferromagnetic ordering. For the cubic case, $a$ is the
    lattice constant and $\Delta_0$ is the fundamental gap. For the tetragonal case, $a$ is the in-plane
    lattice constant, $c/a$ is the ratio of out-of-plane lattice
    constant over in-plane lattice constant,
    $\delta_{\textrm{\textit{B}O}}$ is the $B$-site metal and oxygen displacement along the
    [001] direction. $\Delta_0$ is the fundamental gap and
    $\Delta_{\textrm{optical}}$ is the optical gap. $\Delta E$
    is the energy difference between the tetragonal structure and the
    cubic structure in the unit of meV per 5-atom formula. $P$ is the
    polarization along the [001] direction. $m$ is the local magnetic moment
    on V and Fe sites.}
\begin{center}
\begin{tabularx}{\textwidth}{c| *3{>{\Centering}X}| *3{>{\Centering}X}}
\hline \hline
  & \multicolumn{3}{c|}{Pb$_2$VFeO$_6$}  &
                                           \multicolumn{3}{c}{PbTiO$_3$}\\
\hline xc &  PBE+$U$+$J$ & LDA+$U$+$J$ & sPBEsol & PBE & LDA & PBEsol \\
\hline magnetic & [001]  & [001] & [001] & nm & nm & nm \\
\hline & \multicolumn{6}{c}{cubic structure} \\
\hline $a$ (\AA) & 3.949  &  3.857 & 3.887 &  3.972  & 3.891  & 3.929 \\
\hline $\Delta_0$ (eV) & 0.60  &  0.41 & metallic &  1.61  &  1.47  &  1.53 \\
\hline & \multicolumn{6}{c}{tetragonal structure} \\
\hline $a$ (\AA) & 3.803  & 3.776 & 3.751 &  3.844  & 3.865  & 3.882 \\
\hline $c/a$     & 1.248  & 1.116 &  1.220 &  1.238  & 1.044  & 1.081 \\
\hline $\delta_{\textrm{$B$O}}$ (\AA) & 0.425 (V) 0.629 (Fe)
                      & 0.281 (V) 0.463 (Fe) & 0.413 (V) 0.601 (Fe) & 0.526  & 0.277 & 0.346 \\
\hline $P$ ($\mu$C/cm$^2$) & 124 & 102 & -- & 125 & 79
                                       &  93 \\
\hline $\Delta_0$ (eV) &  0.42 & 0.38 & 0.26 & 1.88 & 1.49 & 1.60 \\ 
\hline $\Delta_{\textrm{optical}}$ (eV) & 1.83 & 1.83 & 1.88 & 2.86 & 2.48 &
                                                                    2.82
  \\
\hline $\Delta E$ (meV) & -251 & -77 & -239 & -209 & -57 & -79 \\
\hline $m$ ($\mu_B$) & 0.147 (V) 4.004 (Fe) & 0.163 (V)
                                4.002 (Fe)&
                                                                  0.183
                                                                  (V)
 3.674 (Fe) & -- & -- & --\\
\hline\hline
\end{tabularx}
\end{center}
\end{table}

\begin{table}[h!]
  \caption{\label{tab:Sr-ferroelectric} \textbf{Comparison of Sr$_2$VFeO$_6$ and
    SrTiO$_3$.} The results are calculated using the DFT method with 
    different exchange correlation functionals (xc). Antiferrodistortive modes are not taken into account in 
    the calculations.
    `nm' stands for non-magnetic and `[001]' for the [001]
    antiferromagnetic ordering. For the cubic case, $a$ is the
    lattice constant and $\Delta_0$ is the fundamental gap. For the tetragonal case, $a$ is the in-plane
    lattice constant, $c/a$ is the ratio of out-of-plane lattice
    constant over in-plane lattice constant,
    $\delta_{\textrm{\textit{B}O}}$ is the $B$-site metal and oxygen displacement along the
    [001] direction. $\Delta_0$ is the fundamental gap and
    $\Delta_{\textrm{optical}}$ is the optical gap. $\Delta E$
    is the energy difference between the tetragonal structure and the
    cubic structure in the unit of meV per 5-atom formula. $P$ is the
    polarization along the [001] direction. $m$ is the local magnetic moment
    on V and Fe sites.}
\begin{center}
\begin{tabularx}{\textwidth}{c| *3{>{\Centering}X}| *3{>{\Centering}X}}
\hline \hline
  & \multicolumn{3}{c|}{Sr$_2$VFeO$_6$}  &
                                           \multicolumn{3}{c}{SrTiO$_3$}\\
\hline xc &  PBE+$U$+$J$ & LDA+$U$+$J$ & sPBEsol & PBE & LDA & PBEsol \\
\hline magnetic &  [001]  & [001] & [001] & nm & nm & nm \\
\hline & \multicolumn{6}{c}{cubic structure} \\

\hline $a$ (\AA) & 3.915  &  3.823  &  3.853  &  3.944  & 3.863  & 3.903 \\
\hline $\Delta_0$ (eV) & 0.40 &  metallic & metallic & 1.79  &  1.80  & 1.81 \\
\hline & \multicolumn{6}{c}{tetragonal structure} \\

\hline $a$ (\AA) & 3.904  & -- & 3.841 &  3.936  & --  & -- \\
\hline $c/a$     & 1.013  & -- &  1.017 &  1.011 & --  & -- \\
\hline $\delta_{\textrm{$B$O}}$ (\AA) & 0.109 (V) 0.120 (Fe)
                      & -- & 0.181 (V) 0.162 (Fe) & 0.120  & -- & -- \\
\hline $P$ ($\mu$C/cm$^2$) & 26 & -- & metallic & 30 & --
                                       &  -- \\
\hline $\Delta_0$ (eV) &  0.30 & -- & metallic & 1.82 & --  & -- \\ 
\hline $\Delta_{\textrm{optical}}$ (eV) & 1.36 & -- & metallic & 2.34 & -- &
                                                                    --
  \\
\hline $\Delta E$ (meV) & -2 & 0 & -34 & -6 & 0 & 0 \\
\hline $m$ ($\mu_B$) & 0.084 (V) 4.089 (Fe) & 0.061 (V)
                                4.107 (Fe)&
                                                                  0.113
                                                                  (V)
 3.543 (Fe) & -- & -- & --\\
\hline\hline
\end{tabularx}
\end{center}
\end{table}

\newpage
\clearpage

%\bibliography{multiferroic-mott-v8}

\end{document}